\documentclass[12pt,twoside]{article}
\usepackage{graphicx}
\usepackage{mathptmx}
\usepackage{amsmath}
\usepackage{amsfonts}
\usepackage{bm}
\usepackage[round]{natbib}

\RequirePackage{fix-cm}

\DeclareMathOperator{\tr}{tr}

\begin{document}

\title{Saturn rings: fractal structure and random field model\thanks{%
This material is based upon the research partially supported by the NSF
under grant CMMI-1462749.}}

\author{Anatoliy Malyarenko\thanks{M\"{a}lardalen University, Sweden} \and Martin Ostoja-Starzewski\thanks{University of Illinois at Urbana-Champaign, USA}}

\date{\today}

\maketitle

\begin{abstract}
This study is motivated by the observation, based on photographs from the
Cassini mission, that Saturn's rings have a fractal structure in radial
direction. Accordingly, two questions are considered: (1) What Newtonian
mechanics argument in support of that fractal structure is possible? (2)
What kinematics model of such fractal rings can be formulated? Both
challenges are based on taking Saturn's rings' spatial structure as being
statistically stationarity in time and statistically isotropic in space, but
statistically non-stationary in space. An answer to the first challenge is
given through the calculus in non-integer dimensional spaces and basic
mechanics arguments (Tarasov (2006) \textit{Celest. Mech. Dyn. Astron.}
\textbf{94}). The second issue is approached in Section~3 by taking the
random field of angular velocity vector of a rotating particle of the ring
as a random section of a special vector bundle. Using the theory of group
representations, we prove that such a field is completely determined by a
sequence of continuous positive-definite matrix-valued functions defined on
the Cartesian square $F^{2}$ of the radial cross-section $F$ of the rings,
where $F$ is a fat fractal.
\end{abstract}

\section{Introduction}

A recent study of the photographs of Saturn's rings taken during the Cassini
mission has demonstrated their fractal structure \citep{Li2015}. This leads
us to ask these questions:

Q1: What mechanics argument in support of that fractal structure is possible?

Q2: What kinematics model of such fractal rings can be formulated?

These issues are approached from the standpoint of Saturn's rings' spatial
structure having (i) statistical stationarity in time and (ii) statistical
isotropy in space, but (iii) statistical non-stationarity in space. The
reason for (i) is an extremely slow decay of rings relative to the time
scale of orbiting around Saturn. The reason for (ii) is the obviously
circular, albeit disordered and fractal, pattern of rings in the radial
coordinate. The reason for (iii) is the lack of invariance with respect to
arbitrary shifts in Cartesian space which, on the contrary and for example,
holds true for a basic model of turbulent velocity fields. Hence, the model
we develop is one of rotational fields of all the particles, each travelling
on its circular orbit whose radius is dictated by basic orbital mechanics.

The Q1 issue is approached in Section 2 from the standpoint of calculus in
non-integer dimensional space, based on an approach going back to %
\citet{doi:10.1142/S0217979205032656,MR2210182}. We compare total energies
of two rings~--- one of non-fractal and another of fractal structure, both
carrying the same mass~--- and infer that the fractal ring is more likely.
We also compare their angular momenta.

The Q2 issue is approached in Section~3 in the following way. Assume that
the angular velocity vector of a rotating particle is a single realisation
of a random field. Mathematically, the above field is a random section of a
special vector bundle. Using the theory of group representations, we prove
that such a field is completely determined by a sequence of continuous
positive-definite matrix-valued functions $\{\,B_{k}(r,s)\colon k\geq 0\,\}$
with
\begin{equation*}
\sum_{k=0}^{\infty }\tr(B_{k}(r,r))<\infty,
\end{equation*}%
where the real-valued parameters $r$ and $s$ run over the radial
cross-section $F$ of Saturn's rings. To reflect the observed fractal nature
of Saturn's rings, \cite{MR610942} and independently \cite{MR665254}
supposed that the set $F$ is a \emph{fat fractal subset} of the set $\mathbb{%
R}$ of real numbers. The set $F$ itself is not a fractal, because its
Hausdorff dimension is equal to $1$. However, the topological boundary $%
\partial F$ of the set $F$, that is, the set of points $x_{0}$ such that an
arbitrarily small interval $(x_{0}-\varepsilon,x_{0}+\varepsilon)$
intersects with both $F$ and its complement, $\mathbb{R}\setminus F$, is a
fractal. The Hausdorff dimension of $\partial F$ is not an integer number.

\section{Mechanics of fractal rings}

\subsection{Basic considerations}

\begin{figure}[htbp]
\centering
\includegraphics[width=\columnwidth]{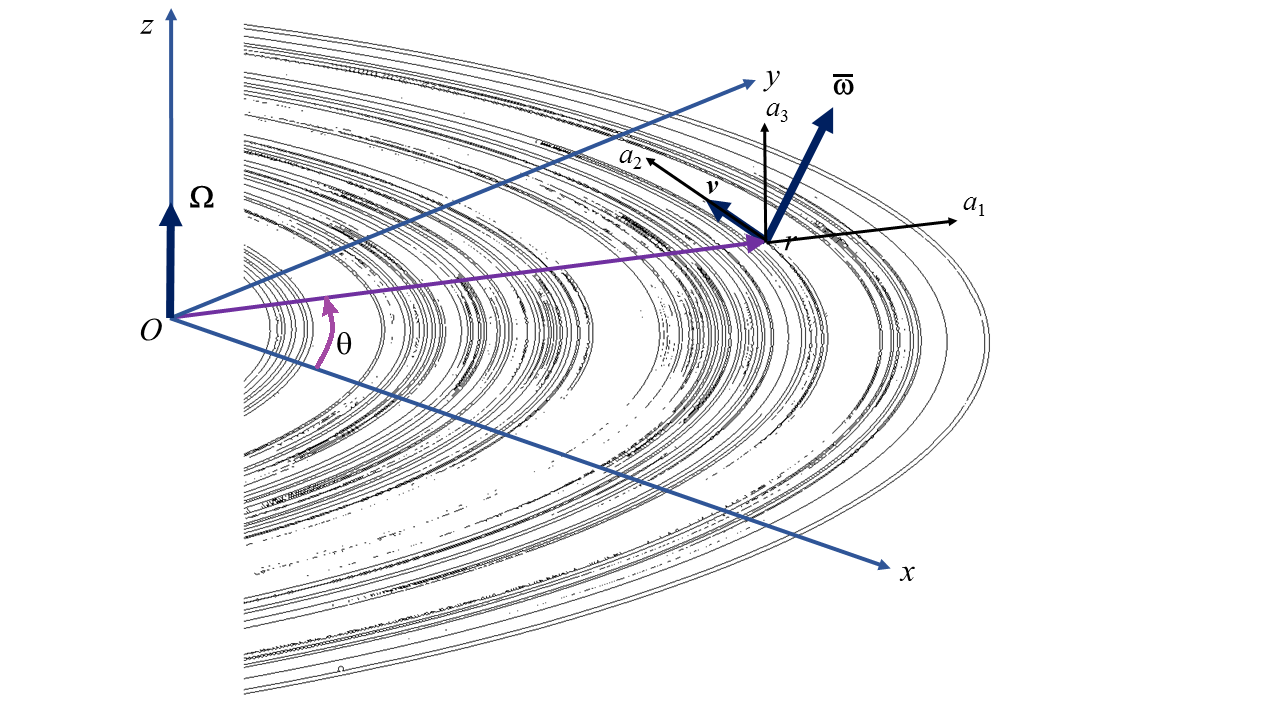}
\caption{The planar ring of particles, adapted from \citep[Fig.~5(b)]{Li2015}, showing the Saturnian (Cartesian and cylindrical)
coordinate systems as well as the orbital frame of reference $\mathbf{(}%
a_{1},a_{2},a_{3})$\ and the  body axes $\mathbf{(}x_{1},x_{2},x_{3})$ of a
typical particle.}
\label{fig:1}
\end{figure}

We begin with the standard gravitational parameter, $\mu =GM_{\mathrm{Saturn}%
}$; its value for Saturn ($\mu =37,931,187$ $km^{3}/s^{2}$) is known but
will not be needed in the derivations that follow.\ For any particle of mass
$m$ located within the ring, we take $m\ll M_{\mathrm{Saturn}}$ with
dimensions also much smaller than the distance to the center of Saturn. Each
particle is regarded as a rigid body, with its orbit about the spherically
symmetric Saturn being circular. We are using the cylindrical coordinate
system $\left( r,\theta ,z\right) $, such that the $z$-axis is aligned with
the normal to the plane of rings, Fig.~\ref{fig:1}. The particle's orbital
frame of reference with the origin $O$ at its center of mass is made of
three axes: $a_{1}$ in the radial direction, $a_{2}$ tangent to the orbit in
the direction of motion, and $a_{3}$\ normal to the orbit plane. All the
particles orbit around Saturn in the same plane. The attitude of any given
particle is described by the vector of body axes $\left\{ \mathbf{x}\right\}
^{T}=\{x_{1},x_{2},x_{3}\}^{T}$, which are related to the vector $\left\{
\mathbf{a}\right\} $ in the orbital frame of reference of the particle\ by%
\begin{equation*}
\left\{ \mathbf{x}\right\} =\left[ \mathbf{l}\right] \left\{ \mathbf{a}%
\right\} .
\end{equation*}%
Here $\left[ \mathbf{l}\right] $\ is the matrix of direction cosines $l_{i}$%
, $i=1,2,3$.

Henceforth, we consider two rings: Euclidean (i.e. non-fractal) and a
fractal one; both rings are planar, Fig.~\ref{fig:1}. Hereinafter the
subscript $_{\mathbb{E}}$ denotes any Euclidean object. Next, we must
consider the mass of a Euclidean ring (body $B_{\mathbb{E}}$) versus a
fractal ring (body $B_{\alpha }$). From a discrete system point of view, the
ring is made of $I$ particles $\{i=1,...,I\}$, each with a respective mass $%
m_{i}$, moment of inertia $\mathbf{j}_{i}$, and positions $\mathbf{x}_{i}$.

The mass of a Euclidean ring $B_{\mathbb{E}}$, with radius $r\in \lbrack
R_{D},R]$ and thickness $h$ in $z$-direction, is now taken in a continuum
sense
\begin{equation}
\begin{array}{c}
M_{\mathbb{E}}=\sum_{i=1}^{I}m_{i}\rightarrow \int_{B_{\mathbb{E}}}\rho dB_{%
\mathbb{E}}=h\int_{R_{D}}^{R}\int_{0}^{2\pi }\rho _{\mathbb{E}}hdS_{2} \\
=2\pi h\rho _{\mathbb{E}}\int_{R_{D}}^{R}rdr=\rho _{\mathbb{E}}h\pi \left(
R^{2}-R_{D}^{2}\right) .%
\end{array}%
\end{equation}%
In the above we have assumed the mass to be homogeneously distributed
throughout the ring with a mass density $\rho _{\mathbb{E}}$. To get
quantitative results, one may take: $R=140\times 10^{6}m$ as the outer
radius of Saturn's F ring, $R_{D}=74.5\times 10^{6}m$ as the radius of the
(inner) D ring, and the rings' thickness $h=100m$.

\subsection{Mass densities}

All the rings constituting the fractal ring $B_{\alpha }$\ are embedded in $%
\mathbb{R}^{3}$, also with radius $r\in \lbrack R_{D},R]$ and thickness $h$
in $z$-direction. The parameter $\alpha $\ ($<1$) denotes the fractal
dimension in the radial direction, i.e. on any ray(any because the ring is
axially symmetric about $z$). Thus, the (planar) fractal dimension, such as
seen and measured on photographs, is $D=\alpha +1<2$, consistent with the
fact that Saturn's rings are partially plane-filling if interpreted as a
planar body. In order to do any analysis involving $B_{\alpha }$, in the
vein of \citet{doi:10.1142/S0217979205032656,MR2210182}, we employ the
integration in non-integer dimensional space. That is, we take the
infinitesimal element $dB_{\alpha }$ of $B_{\alpha }$ according to %
\citep{PhysRevE.88.057001}:
\begin{equation}
dB_{\alpha }=h\text{ }dS_{\alpha },\text{ \ \ }dS_{\alpha }=\alpha \left(
\frac{r}{R}\right) ^{\alpha -1}dS,\text{ \ \ }dS=rdrd\theta .
\end{equation}%
Now, the mass of a fractal ring is
\begin{equation}
\begin{array}{c}
M_{\alpha }=\sum_{i=1}^{I}m_{i}\rightarrow \int_{B}\rho _{\alpha }dB_{\alpha
}=h\int_{R_{D}}^{R}\int_{0}^{2\pi }\rho _{\alpha }dS_{\alpha } \\
=\displaystyle2\pi h\rho _{\alpha }\int_{R_{D}}^{R}\alpha \left( \frac{r}{R}%
\right) ^{\alpha -1}rdr=2\pi h\rho _{\alpha }\frac{\alpha }{\alpha +1}\left(
R^{2}-\frac{R_{D}^{\alpha +1}}{R^{\alpha -1}}\right) ,%
\end{array}%
\end{equation}%
which involves an effective mass density $\rho _{\alpha }$ of a fractal
ring. Note that the above correctly reduces to (1) for $\alpha \rightarrow 1$%
. Since the rings in both interpretations must have the same mass, requiring
$M_{\alpha }=M_{\mathbb{E}}$ for any $\alpha $, gives
\begin{equation}
\rho _{\alpha }=\frac{\alpha +1}{2\alpha }\rho _{\mathbb{E}},
\end{equation}%
which is a decreasing function of $\alpha $\ (i.e. we must have $\rho
_{\alpha }>\rho _{\mathbb{E}}$ for $\alpha <1$) and which correctly gives $%
\lim_{\alpha \rightarrow 1}\rho _{\alpha =1}=\rho _{\mathbb{E}}$ for $\alpha
=1$, i.e. when the fractal ring becomes non-fractal. Thus, a fractal ring
has a higher effective mass density than the homogeneous Euclidean ring of
the same overall dimensions.

\subsection{Moments of inertia}

The moment of inertia of the Euclidean ring ($r\in \lbrack 0,R]$ and
thickness $h$ in $z$-direction), assuming $\rho _{\mathbb{E}}=\mathrm{const}$%
, is%
\begin{equation}
I_{\mathbb{E}}=\frac{1}{2}\pi h\rho _{\mathbb{E}}\left(
R^{4}-R_{D}^{4}\right) =\frac{1}{2}M\left( R^{2}+R_{D}^{2}\right) ,
\end{equation}%
while the moment of inertia of a fractal ring is%
\begin{equation}
\begin{array}{c}
I_{\alpha }=h\int_{B}\rho _{\alpha
}r^{2}dB_{E}=h\int_{R_{D}}^{R}\int_{0}^{2\pi }r^{2}\rho _{\alpha
}hdS_{\alpha } \\
\displaystyle2\pi h\rho _{\alpha }\int_{R_{D}}^{R}r^{2}\alpha \left( \frac{r%
}{R}\right) ^{\alpha -1}rdr=2\pi h\rho _{\alpha }\frac{\alpha }{\alpha +3}%
\left( R^{4}-\frac{R_{D}^{\alpha +3}}{R^{\alpha -1}}\right) .%
\end{array}%
\end{equation}%
Now, take the limit $\alpha \rightarrow 1$:%
\begin{equation}
\lim_{\alpha \rightarrow 1}I_{\alpha }=\frac{1}{2}\pi h\rho _{\mathbb{E}%
}\left( R^{4}-R_{D}^{4}\right) =I_{\mathbb{E}},
\end{equation}%
as expected. Note that $I_{\alpha }$ is an increasing function of $\alpha $\
(i.e. we must have $I_{\alpha }<I_{\mathbb{E}}$ for $\alpha <1$) and which
correctly gives $\lim_{\alpha \rightarrow 1}I_{\alpha }=I_{\mathbb{E}}$ for $%
\alpha =1$. We also observe from (6) that a fractal ring has a lower moment
of inertia than the homogeneous Euclidean ring with the same overall
dimensions.

\subsection{Energies}

Since for an object of mass $m$ on a circular orbit the total energy is $%
E=-\mu /2r$, the total energy (sum of kinetic and potential) of the
Euclidean ring is%
\begin{equation}
\begin{array}{c}
\displaystyle E_{\mathbb{E}}=-\sum_{i=1}^{I}\frac{\mu m_{i}}{2r_{i}}%
\rightarrow -\int_{B}\frac{\mu \rho _{E}}{2r}\text{ }dB \\
=\displaystyle-\frac{1}{2}h\mu \rho _{\mathbb{E}}\int_{0}^{R}\int_{0}^{2\pi
}r^{-1}rdrd\theta =-\pi h\mu \rho _{\mathbb{E}}\left( R-R_{D}\right) .%
\end{array}%
\end{equation}%
On the other hand, the total energy of the fractal ring $B_{\alpha }$ is
[again with $dS_{\alpha }=\alpha \left( \frac{r}{R}\right) ^{\alpha
-1}rdrd\theta $]%
\begin{equation}
\begin{array}{c}
E_{\alpha }=\displaystyle-\sum_{i=1}^{I}\frac{\mu m_{i}}{2r_{i}}\rightarrow
-\int_{B}\frac{\mu \rho _{\alpha }}{2r}\text{ }dB=-\int_{R_{D}}^{R}\frac{1}{2%
}h\mu \rho _{\alpha }\frac{\alpha +1}{2\alpha }r^{-1}dS_{\alpha } \\
=\displaystyle-\int_{R_{D}}^{R}\int_{0}^{2\pi }\frac{1}{2}h\mu \rho _{\alpha
}\alpha \frac{\alpha +1}{2\alpha }r^{-1}\left( \frac{r}{R}\right) ^{\alpha
-1}rdrd\theta =-\pi h\mu \rho _{\alpha }\frac{\alpha +1}{2\alpha }\left(
R-R_{D}\right) .%
\end{array}%
\end{equation}%
Now, take the limit $\alpha \rightarrow 1$:%
\begin{equation}
\lim_{\alpha \rightarrow 1}E_{\alpha }=\frac{1}{2}\pi h\rho _{\mathbb{E}%
}\left( R-R_{D}\right) =E_{\mathbb{E}},
\end{equation}%
as expected.

Comparing $E_{\alpha }$ with $E_{\mathbb{E}}$, gives
\begin{equation}
E_{\alpha }=\frac{\alpha +1}{2\alpha }E_{\mathbb{E}},
\end{equation}%
which is a decreasing function of $\alpha $. Thus, given the minus sign in
(8) and (9), the fractal ring has a lower total energy than the homogeneous
Euclidean ring with the same overall dimensions and the same mass. In other
words, with reference to question Q1 in the Introduction, the ring having a
fractal structure is more likely than that with a non-fractal one.

The foregoing argument extends the approach of \citet{yang2007applied}, who
showed that a Euclidean ring has a lower energy than a Euclidean spherical
shell, which in turn is lower than that of a Euclidean ball. Putting all the
inequalities together, we have%
\begin{equation*}
E_{\alpha }\leq E_{\mathbb{E}}\leq E_{\mathrm{shell}}\leq E_{\mathrm{ball}}.
\end{equation*}

\subsection{Angular Momenta}

For any particle of velocity $v$ on a circular orbit of radius $r$ around a
planet:%
\begin{equation}
\mu =rv^{2}=r^{3}\Omega ^{2}=4\pi ^{2}r^{3}/T^{2},
\end{equation}%
where $\Omega $\ is the angular velocity and $T$ is the period. This
implies:
\begin{equation}
v=\sqrt{\mu /r}\text{\ \ \ and \ \ }\Omega =\sqrt{\mu /r^{3}}.
\end{equation}

For the Euclidean ring ($r\in \lbrack 0,R]$ and thickness $h$ in $z$%
-direction), the angular momentum is%
\begin{equation}
\begin{array}{c}
\displaystyle H_{\mathbb{E}}=\sum_{i=1}^{I}m_{i}r_{i}v_{i}\rightarrow
h\int_{R_{D}}^{R}\int_{0}^{2\pi }\rho _{\mathbb{E}}rv\text{ }rdrd\theta \\
=\displaystyle h\int_{R_{D}}^{R}\int_{0}^{2\pi }\rho _{\mathbb{E}}r\sqrt{\mu
/r}\text{ }rdrd\theta =2\pi h\rho _{\mathbb{E}}\sqrt{\mu }\frac{2}{5}\left(
R^{5/2}-R_{D}^{5/2}\right) ,%
\end{array}%
\end{equation}%
while for the fractal ring $B_{\alpha }$, the angular momentum is%
\begin{equation}
\begin{array}{c}
H_{\alpha }=\sum_{i=1}^{I}m_{i}r_{i}v_{i}\rightarrow
h\int_{0}^{R}\int_{R_{D}}^{2\pi }\rho (r)rv\text{ }dS_{\alpha }=\displaystyle%
2\pi h\rho _{\alpha }\int_{R_{D}}^{R}r\sqrt{\frac{\mu }{r}}\alpha \left(
\frac{r}{R}\right) ^{\alpha -1}\text{ }rdr \\
=\displaystyle2\pi h\rho _{\alpha }\sqrt{\mu }\frac{\alpha }{\alpha +3/2}%
R^{1-\alpha }\left( R^{3/2+\alpha }-R_{D}^{3/2+\alpha }\right) .%
\end{array}%
\end{equation}%
This correctly reduces to $H_{\mathbb{E}}$ above for $\alpha \rightarrow 1$.

Comparing $H_{\alpha }$ with $H_{\mathbb{E}}$, shows that $H_{\alpha }$ is
an increasing function of $\alpha $ and this correctly gives $\lim_{\alpha
\rightarrow 1}H_{\alpha =1}=H_{\mathbb{E}}$, i.e. the fractal ring has a
lower angular momentum than the homogeneous Euclidean ring with the same
overall dimensions.

At this point, we note that in inelastic collisions the momentum is
conserved (just as in elastic collisions), but the kinetic energy is not as
it is partially converted to other forms of energy. If this argument is
applied to the rings, one may argue that $H_{\alpha }=H_{\mathbb{E}}$\
should hold for any $\alpha $, which can be satisfied by accounting for the
angular momentum of particles due to rotation about their own axes . Thus,
instead of (13), writing $j_{i}$\ for the moment of inertia of the particle $%
i$, we have the contribution of the angular momentum of that rotation in
terms of the Euler angle $\phi $ about the $a_{3}$\ axis:%
\begin{equation}
H_{\mathbb{E}}=\sum_{i=1}^{I}m_{i}r_{i}v_{i}+\sum_{i\in I}j_{i}\omega
_{zi}\rightarrow h\int_{R_{D}}^{R}\int_{0}^{2\pi }\rho _{\mathbb{E}}rv\text{
}rdrd\theta +h\int_{R_{D}}^{R}\int_{0}^{2\pi }j\phi \text{ }rdrd\theta .
\end{equation}%
The first integral can be calculated as before, while in the second one we
could assume $j=const$ although this would still leave the microrotation $%
\omega _{z}$ as an unknown function of $r$. Turning to the fractal ring we
also have two terms%
\begin{equation}
H_{\alpha }=\sum_{i=1}^{I}m_{i}r_{i}v_{i}+\sum_{i\in I}j_{i}\omega
_{zi}\rightarrow h\int_{R_{D}}^{R}\int_{0}^{2\pi }\rho _{E}rv\text{ }%
dS_{\alpha }+h\int_{R_{D}}^{R}\int_{0}^{2\pi }j\phi \text{ }dS_{\alpha },
\end{equation}%
showing that the statistics $\omega _{z}\left( r\right) $\ needs to be
determined. At this point we turn to the question Q2.

\section{A stochastic model of kinematics}

First, we consider the particles in Saturn's rings at a time moment $0$.

Introduce a spherical coordinate system $(r,\varphi ,\theta )$ with origin $%
O $ in the centre of Saturn such that the plane of Saturn's rings
corresponds to the polar angle's value $\theta =\pi /2$. Let $\overline{%
\bm{\omega}}(r,\varphi )\in \mathbb{R}^{3}$ be the angular velocity vector
of a rotating particle located at $(r,\varphi )$. We assume that $\overline{%
\bm{\omega}}(r,\varphi )$ is a \emph{single realisation of a random field}.

To explain the exact meaning of this construction, we proceed as follows.
Let $(x,y,z)$ be a Cartesian coordinate system with origin in the centre of
Saturn such that the plane of Saturn's rings corresponds to the $xy$-plane,
Fig.~\ref{fig:1}. Let $O(2)$ be the group of real orthogonal $2\times 2$
matrices, and let $SO(2)$ be its subgroup consisting of matrices with
determinant equal to $1$. Put $G=O(2)\times SO(2)$, $K=O(2)$. The
homogeneous space $C=G/K=SO(2)$ can be identified with a circle, the
trajectory of a particle inside rings.

Consider the real orthogonal representation $U$ of the group $O(2)$ in $%
\mathbb{R}^3$ defined by
\begin{equation}  \label{eq:1}
g=
\begin{pmatrix}
g_{11} & g_{12} \\
g_{21} & g_{22}%
\end{pmatrix}
\mapsto U(g)=
\begin{pmatrix}
g_{11} & g_{12} & 0 \\
g_{21} & g_{22} & 0 \\
0 & 0 & \det g%
\end{pmatrix}
.
\end{equation}
Introduce an equivalence relation in the Cartesian product $G\times\mathbb{R}%
^3$: two elements $(g_1,\mathbf{x}_1)$ and $(g_2,\mathbf{x}_2)$ are
equivalent if and only if there exists an element $g\in O(2)$ such that $%
(g_2,\mathbf{x}_2)=(g_1g,U(g^{-1})\mathbf{x}_1)$. The \emph{projection map}
maps an element $(g,\mathbf{x})\in G\times\mathbb{R}^3$ to its equivalence
class and defines the quotient topology on the set $E_U$ of equivalence
classes. Another projection map,
\begin{equation*}
\pi\colon E_U\to C,\qquad\pi(g,\mathbf{x})=gK,
\end{equation*}
determines a \emph{vector bundle} $\xi=(E_U,\pi,C)$.

The topological space $R=\mathbb{R}^2\setminus\{\mathbf{0}\}$ is the union
of circles $C_r$ of radiuses $r>0$. Every circle determines the vector
bundle $\xi_r=(E_{Ur},\pi_r,C_r)$. Consider the vector bundle $%
\eta=(E,\pi,R) $, where $E$ is the union of all $E_{Ur}$, and the
restriction of the projection map $\pi$ to $E_{Ur}$ is equal to $\pi_r$. The
random field $\overline{\bm{\omega}}(r,\varphi)$ is a \emph{random section}
of the above bundle, that is, $\overline{\bm{\omega}}(r,\varphi)\in%
\pi^{-1}(r,\varphi)=\mathbb{R}^3$. In what follow we assume that the random
field $\overline{\bm{\omega}}(r,\varphi)$ is \emph{second-order}, i.e., $%
\mathsf{E}[\|\overline{\bm{\omega}}(r,\varphi)\|^2]<\infty$ for all $%
(r,\varphi)\in R$.

There are at least three different (but most probably equivalent) approaches
to the construction of random sections of vector bundles, the first by %
\citet{MR2737761}, the second by \citet{MR2884225,MR2977490}, and the third
by \citet{MR3170229}. In what follows, we will use the second named
approach. It is based on the following fact: the vector bundle $%
\eta=(E,\pi,R)$ is \emph{homogeneous} or \emph{equivariant}. In other words,
the action of the group $O(2)$ on the bundle base $R$ induces the action of $%
O(2)$ on the total space $E$ by $(g_0,\mathbf{x})\mapsto(gg_0,\mathbf{x})$.
This action identifies the spaces $\pi^{-1}(r_0,\varphi)$ for all $%
\varphi\in[0,2\pi)$, while the action of the multiplicative group $\mathbb{R}%
^+$ on R, $\lambda(r,\varphi)=(\lambda r,\varphi)$, $\lambda>0$, identifies
the spaces $\pi^{-1}(r,\varphi_0)$ for all $r>0$. We suppose that the random
field $\overline{\bm{\omega}}(r,\varphi)$ is \emph{mean-square continuous},
i.e.,
\begin{equation*}
\lim_{\|\mathbf{x}-\mathbf{x}_0\|\to 0}\mathsf{E}[\|\overline{\bm{\omega}}(%
\mathbf{x})-\overline{\bm{\omega}}(\mathbf{x}_0))\|^2]=0
\end{equation*}
for all $\mathbf{x}_0\in R$.

Let $\langle \overline{\bm{\omega}}(\mathbf{x})\rangle =\mathsf{E}[\overline{%
\bm{\omega}}(\mathbf{x})]$ be the one-point correlation vector of the random
field $\overline{\bm{\omega}}(\mathbf{x})$. On the one hand, under rotation
and/or reflection $g\in O(2)$ the point $\mathbf{x}$ becomes the point $g%
\mathbf{x}$. Evidently, the axial vector $\overline{\bm{\omega}}(\mathbf{x})$
transforms according to the representation \eqref{eq:1} and becomes $U(g)%
\overline{\bm{\omega}}(g\mathbf{x})$. The one-point correlation vector of
the so transformed random field remains the same, i.e.,
\begin{equation*}
\langle \overline{\bm{\omega}}(g\mathbf{x})\rangle =U(g)\langle \overline{%
\bm{\omega}}(\mathbf{x})\rangle .
\end{equation*}%
On the other hand, the one-point correlation vector of the random field $%
\overline{\bm{\omega}}(r,\varphi )$ should be independent upon an arbitrary
choice of the $x$- and $y$-axes of the Cartesian coordinate systems, i.e.,
it should not depend on $\varphi $. Then we have
\begin{equation*}
\langle \overline{\bm{\omega}}(\mathbf{x})\rangle =U(g)\langle \overline{%
\bm{\omega}}(\mathbf{x})\rangle
\end{equation*}%
for all $g\in O(2)$, i.e., $\langle \overline{\bm{\omega}}(\mathbf{x}%
)\rangle $ belongs to a subspace of $\mathbb{R}^{3}$ where a trivial
component of $U$ acts. Then we obtain $\langle \overline{\bm{\omega}}(%
\mathbf{x})\rangle =\mathbf{0}$, because $U$ does not contain trivial
components.

Similarly, let $\langle\overline{\bm{\omega}}(\mathbf{x}),\overline{%
\bm{\omega}}(\mathbf{y})\rangle=\mathsf{E}[\overline{\bm{\omega}}(\mathbf{x}%
) \otimes\overline{\bm{\omega}}(\mathbf{y})]$ be the two-point correlation
tensor of the random field $\overline{\bm{\omega}}(\mathbf{x})$. Under the
action of $O(2)$ we should have
\begin{equation*}
\langle\overline{\bm{\omega}}(g\mathbf{x}),\overline{\bm{\omega}}(g\mathbf{y}%
)\rangle =(U\otimes U)(g)\langle\overline{\bm{\omega}}(\mathbf{x}),\overline{%
\bm{\omega}}(\mathbf{y})\rangle.
\end{equation*}
In other words, the random field $\overline{\bm{\omega}}(\mathbf{x})$ is
\emph{wide-sense isotropic} with respect to the group $O(2)$ and its
representation $U$.

Consider the restriction of the field $\overline{\bm{\omega}}(\mathbf{x})$
to a circle $C_r$, $r>0$. The spectral expansion of the field $\{\,\overline{%
\bm{\omega}}(r,\varphi)\colon\varphi\in C_r\,\}$ can be calculated using %
\citet[Theorem~2]{MR2884225} or \citet[Theorem~2.28]{MR2977490}.

The representation $U$ is the direct sum of the two irreducible
representations $\lambda_-(g)=\det g$ and $\lambda_1(g)=g$. The vector
bundle $\eta$ is the direct sum of the vector bundles $\eta_-$ and $\eta_1$,
where the bundle $\eta_-$ (resp. $\eta_1$) is generated by the
representation $\lambda_-$ (resp. $\lambda_1$). Let $\mu_0$ be the trivial
representation of the group $SO(2)$, and let $\mu_k$ be the representation
\begin{equation*}
\mu_k(\varphi)=
\begin{pmatrix}
\cos(k\varphi) & \sin(k\varphi) \\
-\sin(k\varphi) & \cos(k\varphi)%
\end{pmatrix}
.
\end{equation*}
The representations $\lambda_-\otimes\mu_k$, $k\geq 0$ are all irreducible
orthogonal representations of the group $G=O(2)\times SO(2)$ that contain $%
\lambda_-$ after restriction to $O(2)$. The representations $%
\lambda_1\otimes\mu_k$, $k\geq 0$ are all irreducible orthogonal
representations of the group $G=O(2)\times SO(2)$ that contain $\lambda_1$
after restriction to $O(2)$. The matrix entries of $\mu_0$ and of the second
column of $\mu_k$ form an orthogonal basis in the Hilbert space $L^2(SO(2),%
\mathrm{d}\varphi)$. Their multiples
\begin{equation*}
e_k(\varphi)=
\begin{cases}
\frac{1}{\sqrt{2\pi}}, & \mbox{if } k=0, \\
\frac{1}{\sqrt{\pi}}\cos(k\varphi), & \mbox{if } k\leq-1 \\
\frac{1}{\sqrt{\pi}}\sin(k\varphi), & \mbox{if } k\geq 1%
\end{cases}%
\end{equation*}
form an orthonormal basis of the above space. Then we have
\begin{equation}  \label{eq:3}
\overline{\bm{\omega}}(r,\varphi)=\sum_{k=-\infty}^{\infty}e_k(\varphi)%
\mathbf{Z}^k(r),
\end{equation}
where $\{\,\mathbf{Z}^k(r)\colon k\in\mathbb{Z}\,\}$ is a sequence of
centred stochastic processes with
\begin{equation*}
\begin{aligned}
\mathsf{E}[\mathbf{Z}^k(r)\otimes\mathbf{Z}^l(r)]&=\delta_{kl}B^{(k)}(r),\\
\sum_{k\in\mathbb{Z}}\tr(B^{(k)}(r))&<\infty. \end{aligned}
\end{equation*}

It follows that
\begin{equation*}
\mathbf{Z}^k(r)=\int_{0}^{2\pi}\overline{\bm{\omega}}(r,\varphi)e_k(\varphi)%
\,\mathrm{d}\varphi.
\end{equation*}
Then we have
\begin{equation}  \label{eq:2}
\mathsf{E}[\mathbf{Z}^k(r)\otimes\mathbf{Z}^l(s)]=\int_{0}^{2\pi}\int_{0}^{2%
\pi} \mathsf{E}[\overline{\bm{\omega}}(r,\varphi_1)\otimes\overline{%
\bm{\omega}}(s,\varphi_2)] e_k(\varphi_1)\,\mathrm{d}\varphi_1e_l(\varphi_2)%
\,\mathrm{d}\varphi_2.
\end{equation}
The field is isotropic and mean-square continuous, therefore
\begin{equation*}
\mathsf{E}[\overline{\bm{\omega}}(r,\varphi_1)\otimes\overline{\bm{\omega}}%
(s,\varphi_2)] =B(r,s,\cos(\varphi_1-\varphi_2))
\end{equation*}
is a continuous function. Note that $e_k(\varphi)$ are spherical harmonics
of degree $|k|$. Denote by $\mathbf{x\cdot y}$ the standard inner product in
the space $\mathbb{R}^d$, and by $\mathrm{d}\omega(\mathbf{y})$ the Lebesgue
measure on the unit sphere $S^{d-1}=\{\,\mathbf{x}\in\mathbb{R}^d\colon\|%
\mathbf{x}\|=1\,\}$. Then
\begin{equation*}
\int_{S^{d-1}}\,\mathrm{d}\omega(\mathbf{x})=\omega_d=\frac{2\pi^{d/2}}{%
\Gamma(d/2)},
\end{equation*}
where $\Gamma$ is the Gamma function.

Now we use the Funk--Hecke theorem, see \citet{MR1688958}. For any
continuous function $f$ on the interval $[-1,1]$ and for any spherical
harmonic $S_k(\mathbf{y})$ of degree $k$ we have
\begin{equation*}
\int_{S^{d-1}}f(\mathbf{x\cdot y})S_k(\mathbf{x})\,\mathrm{d}\omega(\mathbf{x%
})=\lambda_kS_k(\mathbf{y}),
\end{equation*}
where
\begin{equation*}
\lambda_k=\omega_{d-1}\int_{-1}^{1}f(u)\frac{C^{(n-2)/2}_k(u)}{%
C^{(n-2)/2}_k(1)} (1-u^2)^{(n-3)/2}\,\mathrm{d}u,
\end{equation*}
$d\geq 3$, and $C^{(n-2)/2}_k(u)$ are Gegenbauer polynomials. To see how
this theorem looks like when $d=2$, we perform a limit transition as $%
n\downarrow 2$. By \citet[Equation~6.4.13']{MR1688958},
\begin{equation*}
\lim_{\lambda\to 0}\frac{C^{\lambda}_k(u)}{C^{\lambda}_k(1)}=T_k(u),
\end{equation*}
where $T_k(u)$ are Chebyshev polynomials of the first kind. We have $%
\omega_1=2$, $\mathbf{x\cdot y}$ becomes $\cos(\varphi_1-\varphi_2)$, and $%
\mathrm{d}\omega(\mathbf{x})$ becomes $\mathrm{d}\varphi_1$. We obtain
\begin{equation*}
\int_{0}^{2\pi}B(r,s,\cos(\varphi_1-\varphi_2))e_k(\varphi_1)\,\mathrm{d}%
\varphi_1 =B^{(k)}(r,s)e_k(\varphi_2),
\end{equation*}
where
\begin{equation*}
B^{(k)}(r,s)=2\int_{-1}^{1}B(r,s,u)T_{|k|}(u)(1-u^2)^{-1/2}\,\mathrm{d}u,
\end{equation*}
Equation~\eqref{eq:2} becomes
\begin{equation*}
\mathsf{E}[\mathbf{Z}^k(r)\otimes\mathbf{Z}^l(s)]=\int_{0}^{2%
\pi}B^{(k)}(r,s) e_k(\varphi_2)e_l(\varphi_2)\,\mathrm{d}\varphi_2=%
\delta_{kl}B^{(k)}(r,s).
\end{equation*}
In particular, if $k\neq l$, then the processes $\mathbf{Z}^k(r)$ and $%
\mathbf{Z}^l(r)$ are uncorrelated.

Calculate the two-point correlation tensor of the random field $\overline{%
\bm{\omega}}(r,\varphi)$. We have
\begin{equation}  \label{eq:4}
\begin{aligned}
\mathsf{E}[\overline{\bm{\omega}}(r,\varphi_1)\otimes\overline{\bm{%
\omega}}(s,\varphi_2)]
&=\sum_{k=-\infty}^{\infty}e_k(\varphi_1)e_k(\varphi_2)B^{(k)}(r,s)\\
&=\frac{1}{2\pi}B^{(0)}(r,s)+\frac{1}{\pi}\sum_{k=1}^{\infty}
\cos(k(\varphi_1-\varphi_2))B^{(k)}(r,s). \end{aligned}
\end{equation}

Now we add a time coordinate, $t$, to our considerations. A particle located
at $(r,\varphi)$ at time moment~$t$, was located at $(r,\varphi-\sqrt{GM}%
t/r^{3/2})$ at time moment~$0$. It follows that
\begin{equation*}
\overline{\bm{\omega}}(t,r,\varphi)=\overline{\bm{\omega}}\left(r,\varphi-%
\frac{\sqrt{GM}t}{r^{3/2}}\right),
\end{equation*}
where $G$ is Newton's gravitational constant and $M$ is the mass of Saturn.
Equation~\eqref{eq:3} gives
\begin{equation}  \label{eq:5}
\overline{\bm{\omega}}(t,r,\varphi)=\sum_{k=-\infty}^{\infty}
e_k\left(\varphi-\frac{\sqrt{GM}t}{r^{3/2}}\right)\mathbf{Z}^k(r),
\end{equation}
while Equation~\eqref{eq:4} gives
\begin{equation*}
\begin{aligned}
\mathsf{E}[\overline{\bm{\omega}}(t_1,r,\varphi_1)\otimes\overline{\bm{%
\omega}}(t_2,s,\varphi_2)] &=\frac{1}{2\pi}B^{(0)}(r,s)\\
&\quad+\frac{1}{\pi}\sum_{k=1}^{\infty}
\cos\left(k\left(\varphi_1-\varphi_2-\frac{\sqrt{GM}(t_1-t_2)}{r^{3/2}}%
\right)\right)B^{(k)}(r,s). \end{aligned}
\end{equation*}

Conversely, let $\{\,B^{(k)}(r,s)\colon k\geq 0\,\}$ be a sequence of
continuous positive-definite matrix-valued functions with
\begin{equation}  \label{eq:6}
\sum_{k=0}^{\infty}\tr(B^{(k)}(r,r))<\infty,
\end{equation}
and let $\{\,\mathbf{Z}_k(r)\colon k\in\mathbb{Z}\,\}$ be a sequence of
uncorrelated centred stochastic processes with
\begin{equation*}
\mathsf{E}[\mathbf{Z}^k(r)\otimes\mathbf{Z}^l(s)]=\delta_{kl}B^{(|k|)}(r,s).
\end{equation*}
The random field \eqref{eq:5} may describe rotating particles inside
Saturn's rings, if all the functions $B^{(k)}(r,s)$ are equal to $0$ outside
the rectangle $[R_0,R_1]^2$, where $R_0$ (resp. $R_1$) is the inner (resp.
outer) radius of Saturn's rings.

To make our model more realistic, we assume that all the functions $%
B^{(k)}(r,s)$ are equal to $0$ outside the Cartesian square $F^{2}$, where $%
F $ is a \emph{fat fractal} subset of the interval $[R_{0},R_{1}]$, see %
\citet{PhysRevLett.55.661}. \citet{MR665254} calls these sets \emph{dusts of
positive measure}. Such a set has a positive Lebesgue measure, its Hausdorff
dimension is equal to $1$, but the Hausdorff dimension of its boundary is
not an integer number.

A classical example of a fat fractal is a \emph{fat Cantor set}. In contrast
to the ordinary Cantor set, where we delete the middle one-third of each
interval at each step, this time we delete the middle $3^{-n}$th part of
each interval at the $n$th step.

To construct an example, consider an arbitrary sequence of continuous
positive-definite matrix-valued functions $\{\,B^{(k)}(r,s)\colon k\geq
0\,\} $ satisfying \eqref{eq:6} of the following form:
\begin{equation*}
B^{(k)}(r,s)=\sum_{i\in I_k}\mathbf{f}_{ik}(r)\mathbf{f}^{\top}_{ik}(s),
\end{equation*}
where $\mathbf{f}_{ik}(r)\colon[R_0,R_1]\to\mathbb{R}^3$ are continuous
functions, satisfying the following condition: for each $r\in[R_0,R_1]$ the
set $I_{kr}=\{\,i\in I_k\colon f_i(r)\neq 0\,\}$ is as most countable and
the series
\begin{equation*}
\sum_{i\in I_{kr}}\|\mathbf{f}_i(r)\|^2
\end{equation*}
converges. The so defined function is obviously positive-definite. Put
\begin{equation*}
\tilde{B}^{(k)}(r,s)=\sum_{i\in I_k}\tilde{\mathbf{f}}_{ik}(r)\tilde{\mathbf{%
f}}^{\top}_{ik}(s),\qquad r,s\in F.
\end{equation*}
The functions $\tilde{B}^{(k)}(r,s)$ are the restrictions of
positive-definite functions $B^{(k)}(r,s)$ to $F^2$ and are
positive-definite themselves. Consider the centred stochastic process $\{\,%
\tilde{\mathbf{Z}}^k(r)\colon r\in F\,\}$ with
\begin{equation*}
\mathsf{E}[\tilde{\mathbf{Z}}^k(r)\otimes\tilde{\mathbf{Z}}^l(s)]
=\delta_{kl}\tilde{B}^{(|k|)}(r,s),\qquad r,s\in F.
\end{equation*}
Condition \eqref{eq:6} guarantees the mean-square convergence of the series
\begin{equation*}
\overline{\bm{\omega}}(t,r,\varphi)=\sum_{k=-\infty}^{\infty}
e_k\left(\varphi-\frac{\sqrt{GM}t}{r^{3/2}}\right)\tilde{\mathbf{Z}}^k(r)
\end{equation*}
for all $t\geq 0$, $r\in F$, and $\varphi\in[0,2\pi]$.

\section{Closure}

This paper reports an investigation of the fractal character of Saturnian
rings. First, working with the calculus in a non-integer dimensional space,
by energy arguments, we infer that the fractally structured ring is more
likely than a non-fractal one. Next, we develop a kinematics model in which
angular velocities of particles form a random field.

\bibliographystyle{abbrvnat}
\bibliography{Saturn}

%
%

\end{document}